\documentclass[prl,aps,showpacs,twocolumn,unsortedaddress,preprintnumbers]{revtex4}
\usepackage{amsfonts,amsmath,amssymb,ragged2e,comment}
\usepackage[dvips]{graphicx}

\usepackage[margin=0pt,font=small,skip=0pt]{caption}
\let\Oldcaptionfont\captionfont
\renewcommand{\captionfont}{\Oldcaptionfont\justifying} 
\newcommand{\hBS}{\mathbf{\hat S}}
\newcommand{\BB}{\mathbf{B}}
\newcommand{\Bm}{\mathbf{m}}
\newcommand{\Bn}{\mathbf{n}}
\newcommand{\Bk}{\mathbf{k}}
\newcommand{\Bx}{\mathbf{x}}

\renewcommand{\[}{\left[}
\renewcommand{\]}{\right]}

\DeclareMathOperator{\arctanh}{{\rm arctanh}}

\begin{document}

\preprint{ITP-UU-09/20}

\title{The imbalanced antiferromagnet in an optical lattice}
\author{Arnaud Koetsier}
\email{a.o.koetsier@uu.nl}
\author{F. van Liere}
\author{H.~T.~C. Stoof}
\affiliation{Institute for Theoretical Physics, Utrecht University, Leuvenlaan 4, 3584 CE Utrecht, The Netherlands}

\date{\today}


\begin{abstract}
We study the rich properties of the imbalanced antiferromagnet in an optical lattice. We present its phase diagram, discuss spin waves and explore the emergence of topological excitations in two dimensions, known as merons, which are responsible for a Kosterlitz-Thouless transition that has never unambiguously been observed.
\end{abstract}

\maketitle

\textit{Introduction.} --- Owing to their exquisite experimental tunability, ultracold atomic gases have become a test bed for many paradigmatic ideas in quantum many-body physics. In particular, ultracold atoms confined in an optical lattice, i.e., a periodic potential for neutral atoms created by orthogonal retroreflected laser beams, are accurately described by the Hubbard model. The narrow energy bands resulting from the periodic potential quench the kinetic energy of the atoms with respect to their interaction energy, enabling the exploration of strongly correlated phases that play a significant role in condensed-matter physics. The single-band Hubbard model is realised by cold atoms when the lattice potential is sufficiently strong that only the lowest-energy band is populated~\cite{jaksch98}. At a filling corresponding to one particle per lattice site, this model has a Mott-insulator phase for repulsive interactions, and in an atomic Bose gas the theoretically predicted superfluid-to-Mott-insulator phase transition~\cite{fischer1989} has indeed been observed experimentally~\cite{greiner2002}. For bosons, one commonly refers to this model as the Bose-Hubbard model. The Fermi-Hubbard model, referred to simply as the Hubbard model, has also been realised experimentally~\cite{kohl2004}, and the fermionic Mott-insulator has only recently been seen~\cite{jordans2008,schneider2008}.

Aside from being rich in novel physics in itself, this latter system may also shed new light on the poorly understood phenomenon of high-temperature superconductivity in the cuprates. There, the Hubbard model is thought to describe electrons in the periodic ion-lattice potential of the copper-oxygen planes which are believed to undergo a quantum phase transition to a $d$-wave superconducting state as the filling fraction is reduced by doping~\cite{LeeNagaosaWen2006}. In this case, electrons on the same lattice site repel each other due to the Coulomb interaction. We therefore consider here the repulsive Hubbard model, although the attractive Hubbard model can also be realised and possesses interesting BEC-BCS crossover physics~\cite{koetsier2006}.

While achieving the antiferromagnetic ground state of the Hubbard model in an atomic Fermi gas is currently a major experimental goal, another prominent topic that is the focus of a number of recent ground-breaking experiments is imbalance in two-component Fermi gases~\cite{zwierlein2006,Partridge2006}. There, the population of each spin species can be controlled permitting the exploration of uncharted imbalanced phases of great interest in condensed-matter, nuclear, high-energy, and astroparticle physics. 
In this Letter, we explore the bridge between these two exciting directions in the physics of ultracold atomic gases.

\begin{figure}
\begin{center}
 \includegraphics[width=.72\columnwidth]{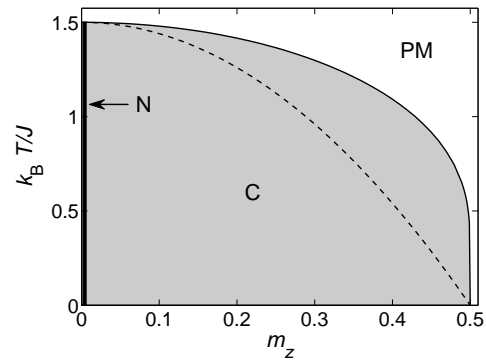} 
\end{center}
\caption{Phase diagram of the imbalanced antiferromagnet in three dimensions. The solid curve is the critical temperature $k_B T / J$ for the canted antiferromagnet (C). Above the critical temperature the system is paramagnetic (PM). Here, $J$ is the superexchange coupling of the antiferromagnetic Heisenberg model.
Also shown is the critical temperature for the corresponding Ising model (dashed line).
In the absence of imbalance the system is in a pure antiferromagnetic or N\'eel (N) state.}
\label{phasediag}
\end{figure}

\textit{Phase diagram} --- In the Mott-insulator phase only spin degrees of freedom remain and its low-lying excitations are then effectively described by the Heisenberg model. The three-dimensional phase diagram calculated within mean-field theory for this model is shown in Fig.~\ref{phasediag} for a cubic lattice which is experimentally most relevant.
When there are equal proportions of each spin species, the Mott insulator becomes a pure antiferromagnetic or N\'eel state below a certain critical temperature and is solely characterised by a nonzero expectation value of the N\'eel order parameter vector $\Bn$, also known as the staggered magnetisation. Changing the proportion of each spin species gives rise to a nonzero average magnetisation  $\Bm=(0,0,m_z)$, where $m_z=S(N_{\uparrow} - N_{\downarrow})/(N_{\uparrow} + N_{\downarrow})$ is equal to the average spin per site of the system. Here, $S=1/2$ is the atomic pseudospin for the two-component Fermi mixture of interest, and $N_{\uparrow,\downarrow}$ denotes the number of particles of each spin species. In this imbalanced case, the system has a canted antiferromagnetic phase at low temperatures where the N\'eel vector is always found to be perpendicular to the magnetisation~\cite{vdongen09}. Imbalance therefore breaks the rotational symmetry of the system and defines an easy plane for the N\'eel vector that is perpendicular to the average magnetisation. 

\begin{figure}[t]
\begin{center}
\includegraphics[width=.75\columnwidth]{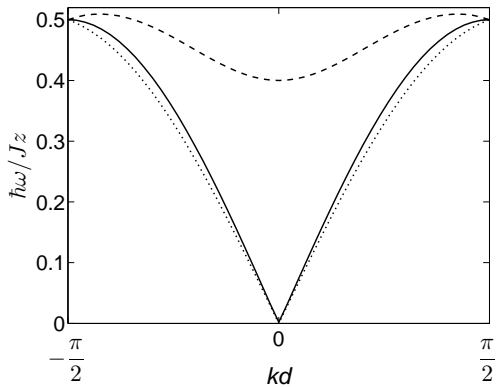}
\end{center}
\caption{The dispersion relation for the spin waves. Here, $\Bk=k(1,1,1)$ is the wave vector and $d$ is the lattice spacing. For zero magnetisation (solid line), the dispersion is the usual doubly degenerate gapless antiferromagnetic dispersion. Imbalance splits the degeneracy and one of the dispersions (dashed line) becomes ferromagnetic and 
acquires a gap equal to $2Jzm_z$ where $z$ is the number of nearest neighbours, while the other remains antiferromagnetic (dotted line). The imbalanced dispersions shown here are for $m_z=0.2$.}
\label{modes}
\end{figure}

In both the N\'eel and canted phases the low-energy excitations are spin-density waves called magnons. By linearising the Heisenberg equations of motion of the on-site spin operators, we find that for the balanced situation the magnons have the usual doubly degenerate antiferromagnetic dispersion that is linear for small momenta, as shown in Fig.~\ref{modes}. However, in the imbalanced case the nonzero average magnetisation lifts this degeneracy. As a result, one of the dispersions becomes gapped corresponding to Larmor precession of the spins around the effective magnetic field generated by the nonzero average magnetisation. The gapped dispersion is quadratic at small momenta signalling the appearance of magnons with a ferromagnetic character. 

In experiments the atomic gas is always confined in a trap. However, including a smooth harmonic trapping potential does not lead to inhomogeneities in the Mott-insulator state but places a limit on the total number of particles beyond which the Mott insulator is destroyed~\cite{koetsier08}. As a result, the excitation spectrum shown in Fig.~\ref{modes} becomes discretised in the trap due to finite-size effects. In the Mott-insulator phase, the system in a trap is thus essentially homogeneous and we may reasonably neglect the trap here~\cite{snoek08}. 

\textit{Merons} --- The long-wavelength dynamics obtained above can be summarised by the following non-linear sigma model~\cite{SarmaSachdev98} with an action
\begin{multline}
  S[\Bn(\Bx,t)] = \int \mathrm{d}t \int \frac{\mathrm{d} \Bx}{d^{D}}
  \Biggl\lbrace
   \frac{1}{4Jzn^2}\bigg(\hbar\frac{\partial \Bn(\Bx,t)}{\partial t}
   \\
   -  2Jz\Bm\times\Bn(\Bx,t) \bigg)^2 - \frac{Jd^2}{2}\[\nabla \Bn(\Bx,t)\]^2
  \Biggr\rbrace.
  \label{nlsm}
\end{multline}
Here $z=2D$ is the number of nearest neighbours in the $D$-dimensional hypercubic lattice, $d$~is the lattice spacing, $\Bm=(0,0,m_z)$ is the average magnetisation per site, and $\Bn(\Bx,t)$ the local staggered magnetisation at position~$\Bx$ and time~$t$ with fixed length~$n$. The equilibrium value of the staggered magnetisation is found from minimising the Landau free energy, given by
\begin{equation}
  F[\Bn(\Bx),\Bm]=
  \int \frac{\mathrm{d} \Bx}{d^{D}}
  \left\lbrace
    \frac{Jd^2}{2}\[\nabla \Bn(\Bx)\]^2
  + f[\Bn(\Bx),\Bm]
  \right\rbrace,
\end{equation}
where $f[\Bn(\Bx),\Bm]$ is the on-site free energy which, as we have seen, breaks the rotational symmetry of the system in the presence of imbalance. 
\begin{figure*}
\begin{center}
\begin{tabular}{c@{\hspace*{-.21in}}c}
 \includegraphics[width=1.4\columnwidth]{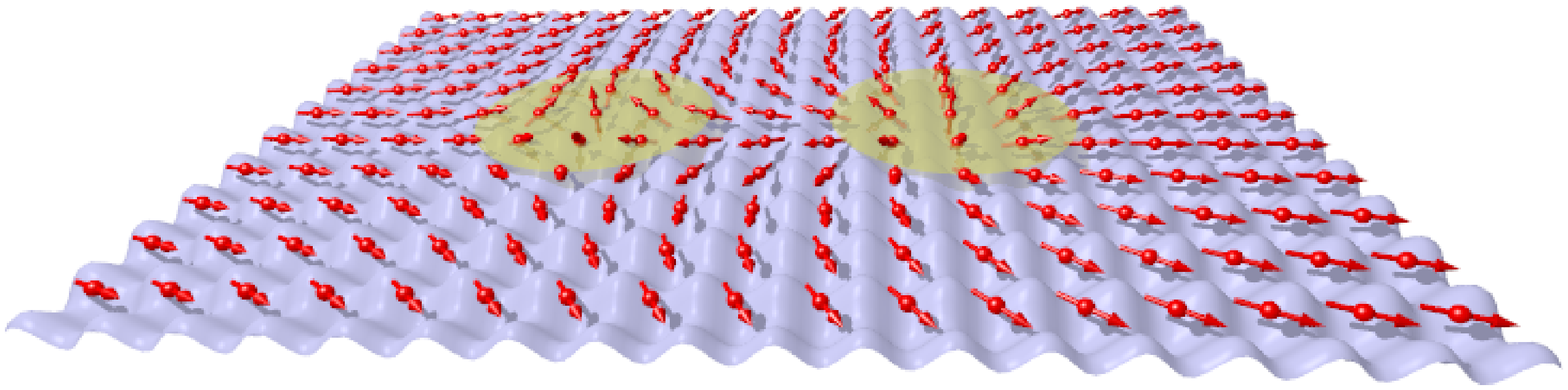}&
 \includegraphics[width=.68\columnwidth]{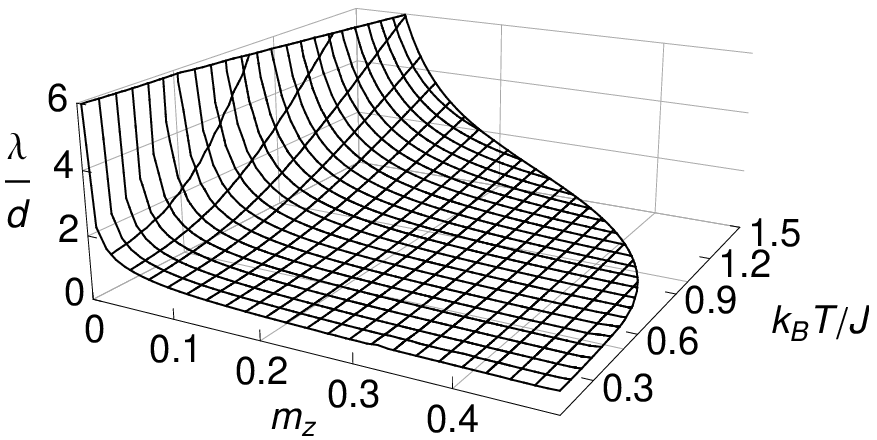}\\
 \textbf{\small (a)}&\textbf{\small (b)}
\end{tabular}
\vspace*{-5pt}
\end{center}
\caption{Merons in two dimensions. \textbf{(a)} A pair of merons with the same Pontryagin index but opposite vorticities. The arrows indicate the staggered magnetization, the periodic optical lattice potential is shown beneath and the core regions of the merons are shown as yellow discs. \textbf{(b)} The radius of the core region $\lambda$ is plotted as a function of temperature and imbalance. The core size diverges as the imbalance is reduced to zero since full rotational symmetry is then restored and merons no longer exist. The surface terminates at the critical temperature where the system becomes paramagnetic.}
\label{meron}
\end{figure*}

\begin{figure}
\begin{center}
\includegraphics[width=.69\columnwidth]{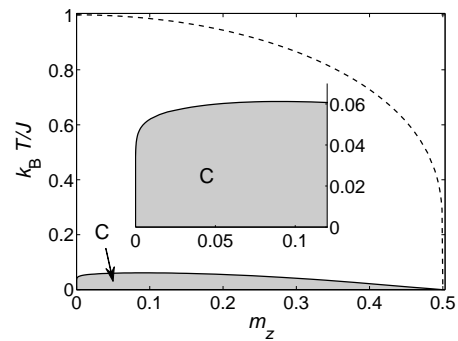}
\end{center}
\caption{Phase diagram of the imbalanced antiferromagnet in two dimensions. The critical temperature for the canted phase (C) is drastically reduced compared to the mean-field result (dashed line) due to fluctuations and becomes a Kosterlitz-Thouless phase transition. Inset: a close-up of the Kosterlitz-Thouless transition for small imbalance.
}
\label{fig:kt}
\end{figure}

The equation of motion that follows from the non-linear sigma model exhibits long-wavelength spin waves discussed above but also admits a class of topologically stable excitations. In the three dimensional case, these are coreless vortices with a spin component near the location of the vortex line that is out of the easy plane. In the two-dimensional case the topological excitations, known as merons~\cite{gross78,affleck86}, are particularly interesting. Far away from the core of the meron, $\Bn(\Bx)$ lies in the easy plane and forms a vortex with vorticity of  $\pm1$, while in the core region $\Bn(\Bx)$ smoothly rotates either up or down out of the easy plane. This spin texture is characterised by two topological invariants, namely, the vorticity and the Pontryagin index which is equal to $\pm1/2$ depending on the polarity of the core spin~\cite{PhysRevB.51.5138}. The Pontryagin index does not play an important role for our purposes since, due to the symmetry of the action under the reflection $\Bn\rightarrow -\Bn$, merons differing only in their Pontryagin index have the same energy. Moreover, the interaction between two merons with a large separation is insensitive to the precise spin configuration in the core. The energy of a single meron diverges logarithmically with the system area $A$ as $(Jn^2\pi/2)\ln(A/\pi\lambda^2)$. Here, $\lambda=d n\sqrt{\pi J/2F_m}$ is the characteristic size of the core, plotted in Fig.~\ref{meron}(b), with $F_m$ the integrated on-site free energy required for the formation of a meron spin texture of size $\lambda=d$. Thus, for large system sizes of interest here, it is impossible to thermally excite a single meron below the critical temperature. This behaviour is analogous to the divergence in the kinetic energy of a single vortex in a two-dimensional Bose-Einstein condensate. However, the energy of a pair consisting of a meron together with another meron of opposite vorticity, or antimeron, is finite since the topological deformation of the spin texture cancels far away from the centre of the pair where the N\'eel vector always points in the same direction, as illustrated in Fig.~\ref{meron}(a).
%
%
Such meron-antimeron pairs are thermally excited below the critical temperature and are responsible for a Kosterlitz-Thouless phase transition associated with the unbinding of the pairs~\cite{kosterlitz73}. 
Indeed, the entropic contribution of the merons above the critical temperature is such that the system can lower its free energy through the proliferation of single merons. Topological excitations are notably absent from the mean-field analysis and as a result our mean-field analysis becomes even qualitatively incorrect in two dimensions. To correct for this, we used Monte Carlo results for a similar system~\cite{klomfass91} to estimate the critical temperature for the Kosterlitz-Thouless transition from the canted antiferromagnet to the paramagnetic phase. We found that this transition occurs at a significantly lower temperature than predicted by mean-field theory in two dimensions, as shown in Fig.~\ref{fig:kt}. In particular, the phase transition occurs at zero temperature for the balanced case where the system rotationally symmetric. In three dimensions, however, we do not expect topological excitations or spin waves to significantly alter the phase diagram. Indeed, numerical studies yield only a 36\% downward correction of the critical temperature in the balanced case~\cite{Staudt2000}. 

\textit{Outlook and conclusions} ---
The imbalanced antiferromagnet can be readily investigated experimentally with modern techniques. Imbalance is achieved in an ultracold Fermi gas by driving spin transitions with an RF field, and a N\'eel-ordered state can be realised by, for example, adiabatically ramping up the optical lattice potential~\cite{koetsier08}. The N\'eel character can then be probed by the measurement of correlations in atom shot noise~\cite{Bloch05,Greiner05} or using Bragg reflection~\cite{fabbri09,clement09} which also enables spin waves to be examined. The unbinding of superfluid vortices in two dimensions has been recently observed in a Bose gas~\cite{hadzibabic06}, and a similar interference experiment could in principle be performed to detect the Kosterlitz-Thouless phase transition here. Moreover, \textit{in situ} imaging techniques based on scanning electron microscopy could be applied to explore topological and magnon excitations by resolving single lattice sites~\cite{ott08,ott09}. Finally, merons possess an internal Ising degree of freedom associated with their Pontryagin index that is very suggestive of their application as a qubit in topological quantum computation, and this is a particularly interesting topic for further investigation. 



\textit{Methods} --- At half filling and for $k_B T \ll U$, with $U$ being the energy cost of adding a second particle to a singly-occupied lattice site, the system is in the Mott-insulator phase and its low-lying excitations are described by the effective antiferromagnetic Heisenberg Hamiltonian
. Imbalance in the number of atoms of each spin can be incorporated by means of a constraint, which enforces that the average total spin is equal to the desired magnetisation. Thus, the Hamiltonian of the imbalanced system is
$\hat H = (J/2)\sum_{\langle i,j\rangle} \hBS_i\cdot \hBS_j - \sum_i{\bf B}\cdot(\hBS_i-\Bm)$,
where $\hbar\hBS_i$ is the spin-$\frac{1}{2}$ operator on site $i$, $\langle i,j\rangle$ denotes a sum over nearest-neighbour sites of the bipartite hypercubic lattice of interest here, and the effective magnetic field $\BB$ acts as a Lagrange multiplier. The exchange constant $J$ arises from the superexchange mechanism and is positive. That is, the system can lower its energy by virtual nearest-neighbour hops only when there is antiferromagnetic ordering. Note that the total spin operator $\sum_i\hBS_i$ commutes with $\hat H$ and thus the magnetisation is a constant of the motion.

Within the usual mean-field analysis we obtain the free energy per site
\begin{multline}
    f(\Bn,\Bm;\BB) = \frac{Jz}{2}(\Bn^2-\Bm^2) +\Bm\cdot\BB \\
    - \frac{1}{2}k_B T\ln\left[4
    \cosh\left(\frac{|\BB_{A}|}{2k_B T}\right)
    \cosh\left(\frac{|\BB_{B}|}{2k_B T}\right)\right],
    \label{f}
\end{multline}
where $\BB_{A\;(B)} = \BB-Jz\Bm \pm Jz\Bn$ and we take the upper (lower) sign for the $A$ ($B$) sublattice, $T$ is the temperature, $k_B$ is the Boltzmann constant, and $\Bn$ is the staggered magnetisation which is related to the average on-site value of the spin by $\langle \hBS_{A (B)} \rangle = \Bm \pm\Bn$. The Lagrange multiplier $\BB$ is then found from the constraint
$\partial f(\Bn,\Bm;\BB) / \partial \BB = \mathbf{0}$.
Minimising the free energy subject to this constraint, we find the critical temperature at which $\langle \Bn \rangle$ becomes nonzero to be $T_c = J z m_z/[2 k_B \arctanh(2 m_z)]$. At this point and throughout the antiferromagnetic phase, the constraint is solved by $\BB=2Jz\Bm$ and $\langle \Bn\rangle$ is perpendicular to $\Bm$ in the minimum of the free energy. By contrast, in the Ising model where $\Bn$ and $\Bm$ are restricted to be parallel, the critical temperature one obtains is $T_c = zJ(1/4 - m_z^2)/k_B$.

We perform a variational calculation substituting the spin-texture of the meron into  Eq.~(\ref{nlsm}) to determine its characteristic size for a range of temperatures and imbalance. A meron spin texture in the $x$-$y$ plane with vorticity $n_v=\pm1$ can be described by
$
\Bn = \lbrace
\sqrt{n^2-[n_z(r)]^2}\cos\phi, n_v\sqrt{n^2-[n_z(r)]^2}\sin\phi, n_z(r)
\rbrace
$,
%
where $n=|\langle \Bn \rangle |$ and $\phi$ is the azimuth in the $x$-$y$ plane. Exact meron solutions that follow from minimising the non-linear sigma model action with this texture have an out-of-plane component that behaves as $n-n_z(r)\propto r^2$ near the origin and that decays exponentially far from the origin \cite{ghosh98}. A suitable variational ansatz that has the right properties near the origin but decays only as a power law for large $r$ is $n_z(r) = n/[(r/\lambda)^2 +1]^2$. In this case, the energy of a single meron due to the gradient term in the non-linear sigma model is $Jn^2\pi[(511/60) - \pi - 6\ln2 + (1/2)\ln(A/\pi\lambda^2) ]+\mathcal{O}(A^{-1})$
which diverges as the logarithm of the area $A$ of the system. Incorporating the correct exponential decay for large distances changes the constant terms in this expression but leaves the logarithm unaffected, indicating that our ansatz is indeed suitable to determine the characteristic core radius $\lambda$. Minimizing the sum of the gradient energy and the integrated on-site free energy in Eq.~(\ref{f}) of the meron texture leads to the optimal size shown in Fig.~\ref{meron}(b). The Pontryagin index here is $+1/2$ and a texture with index $-1/2$ produced by substituting $n_z(r)\rightarrow -n_z(r)$ will produce an identical variational result, as noted earlier. Also, the core structure will play no role in the interaction so long as the spacing of a pair of merons is larger than $\lambda$.

The Kosterlitz-Thouless critical temperature has been calculated for an anisotropic $O(3)$ model using Monte Carlo techniques~\cite{klomfass91}. By making an analogy between the free energy contribution of our non-linear sigma model and the anisotropy in the anisotropic $O(3)$ model, we can use the numerical results of the anisotropic $O(3)$ model to estimate the critical temperature for the imbalanced antiferromagnet that incorporates fluctuations beyond mean-field theory. Although for $m_z \lesssim 0.2$ our anisotropy is more complicated than that used in the Monte Carlo simulations, we do not expect a rigourous calculation to deviate significantly from the results in Fig.~\ref{fig:kt}.


\begin{acknowledgments}
We are very grateful to Rembert Duine and Randy Hulet for providing useful comments and Niels Pannevis for his help with obtaining the results of the imbalanced Ising antiferromagnet. This work is supported by the Stichting voor Fundamenteel Onderzoek der Materie (FOM) and the Nederlandse Organisatie voor Wetenschaplijk Onderzoek (NWO).
\end{acknowledgments}



\end{document}